\documentclass[twocolumn,showpacs,preprintnumbers,amsmath,amssymb]{revtex4}
\usepackage{graphicx}
\usepackage{dcolumn}
\usepackage{bm}

\begin{document}


\title{Bosonization approach to the edge reconstruction of two dimentional electron systems in a quantum dot}

\author{Itaru Yanagi}
\email{yanagi@suou.waseda.jp}
\author{Susumu Kurihara}
\affiliation{Department of Physics, Waseda University, Okubo, Shinjuku, Tokyo 169-8555, Japan}

\date{\today}

\begin{abstract}
We consider the edge reconstruction of electrons in a two dimensional harmonic trap under a strong magnetic field. In this system the edge reconstruction occurs as a result of competition between electron-electron interaction and confining potential. To describe it, we develop a bosonization scheme for two dimensional electron systems. With this method we obtain the excitation spectrum and demonstrate that the edge reconstruction occurs when the value of the magnetic field reaches a critical value. We also show that the edge reconstruction depends on the number of electrons. Additionally, we calculate the third order terms of bosons in Hamiltonian and examine the effect of those terms with a perturbation theory.
\end{abstract}

\maketitle

\section{\label{sec:level1}Introduction}

In recent years technological advances have made it possible to construct ideal two dimensional electron systems in quantum dots~\cite{rf:1}. These devices allow us to study various behaviors of correlated electron systems. 

In this paper, we study the edge reconstruction of two dimensional electron systems in a quantum dot under a strong magnetic field.
 It is a transition caused by interplay between electron-electron interaction and confinement potential.  At relatively strong magnetic fields, all electrons are spin polarized and exist only in the lowest Landau level (LLL). When the magnetic field is below a critical value, electrons in a dot form a compact droplet (filling factor $\nu=1$), occupying orbitals with the lowest total angular momentum. If the magnetic field reaches the critical value, the compact droplet is replaced by the system composed of a smaller droplet surrounded by a ring, with vacancies introduced not into the center, but rather near the droplet edge.  Such a transition has been studied with numerical analyses~\cite{rf:3,rf:4}. Our purpose is to describe it in more analytical form, using bosonization approach. 
 
 There are some other works about bosonization for two dimensional electron systems. Westfahl {\it et al.} introduced Landau level bosonization of two dimentional electron systems~\cite{rf:7}. Rojt {\it et al.} reported Tomonaga-Luttinger (TL) liquid behavior in a two dimensional circular quantum dot at zero magnetic field~\cite{rf:8}.
  
 The edge reconstruction is also studied at fractional quantum Hall edges (FQHE)~\cite{rf:9,rf:10,rf:11,rf:12}. The difference between FQHE and our model is a shape of potential. While the confinement potential of our model is a weak hermonic trap, FQHE is made by sharp boundaries and supposed to form chiral Luttinger liquids~\cite{rf:13}. The mechanism of the reconstruction is essentially the same, i.e., interplay between electron-electron interaction and confinement potential.
 \section{Model and Hamiltonian}
In our model, electrons are confined in a two dimensional harmonic trap under a strong magnetic field perpendicular to the plane,
\begin{eqnarray}
H=\sum_{i=1}^{N}\left(\frac{({\bf p}_{i}-e{\bf A}_{i})^2}{2m^*}+\frac{1}{2}m^*\omega_{0}^2{\bf r}_{i}^2\right)+\sum_{i<j}V(|{\bf r}_{i}-{\bf r}_{j}|)\ , 
\nonumber \\
\end{eqnarray} 
where $N$ is the number of electrons in the quantum dot, $m^*$ is the effective mass of electrons ($m^*=0.067m_{e}$), $\omega_{0}$ is a harmonic oscillator frequency and $V(|{\bf r}|)$ is Coulomb interaction of electrons.
 
 Now all electrons are assumed to be spin-polarized and exist only in the LLL due to a strong magnetic field, the noninteracting part of the Hamiltonian is easily diagonalized and its eigenstates and energies are given by
\begin{eqnarray}
\Psi_{n}(r,\theta)=\frac{1}{l_{0}}\sqrt{\frac{1}{\pi n!}}\left(\frac{r}{l_{0}}\right)^n\exp{\left(-\frac{r^2}{2l_{0}^2}+in\theta\right)}\ ,\\ \nonumber \\
\epsilon_{n}=\left(\hbar\sqrt{\omega_{0}^2+\frac{1}{4}\omega_{c}^2}-\frac{1}{2}\hbar\omega_{c}\right)n+\hbar\sqrt{\omega_{0}^2+\frac{1}{4}\omega_{c}^2}\ ,
\end{eqnarray}
where $n$ is a quantum number of an angular momentum ($n=0,1,2...$), $\omega_{c}$ is a cyclotron frequency, $l_{0}=\sqrt{\hbar/2m^*\Omega}$ and $\Omega=\sqrt{\omega_{0}^2+\omega_{c}^2/4}$. The degeneracy in the LLL is lifted by a parabolic potential. 

\section{Comparison with one dimensional electron systems}
In order to compare this model with one dimensional TL model, we rewrite the Hamiltonian in terms of second quantization,
\begin{eqnarray}
H=\sum_{l}\epsilon_{l}c_{l}^\dagger c_{l}+\frac{1}{2}\sum_{j,k,l}V_{k,j,l}c_{l-k}^\dagger c_{j+k}^\dagger c_{j} c_{l}\ ,
\end{eqnarray}
where $c_{l}^\dagger$ is an operator that creates an electron with angular momentum $l$ in the LLL and $V_{k,j,l}$ is the matrix element of Coulomb interaction.

$\epsilon_{l}$ has the linear dispersion similar to the kinetic energy part in TL model. Our model, however, has only one linear dispersion branch while TL model has two linear dispersion branches, so we don't have to consider the interaction between two different branches. 

$V_{k,j,l}$ has the essential deviation from the TL model because it has not only $k$-dependence but also $j,l$-dependence. In the case of one dimensional electron systems, there is only $k$-dependence which corresponds to the momentum transfer between two particles. 

  If $k$-dependence is much stronger than $j,l$-dependence, it would appear that a bosonization method is effective for our model. $V_{k,j,l}$ can be calculated exactly in this model~\cite{rf:2,rf:3}. The Appendix shows the function form of $V_{i,j,k}$. We evaluate $k$-dependence and $j,l$-dependence of $V_{k,j,l}$, then we find that $k$-dependence is more dominant than $j,l$-dependence ( $V_{k+1,j,l}-V_{k,j,l}>>V_{k,j+1,l}-V_{k,j,l}$, less than 1/10 for several values of $i,j,k$ ).
  
   But then, the edge reconstruction is caused by $j,l$-dependence, so we have to develop a bosonization scheme including $j,l$-dependence.
\section{Bosonization for two deimensional electron systems}
 First, we describe fermionic operators $c$, $c^\dagger$ in terms of bosonic operators $b$, $b^\dagger$. Bosonization of fermionic operators has been studied well~\cite{rf:5}, but this time we cannot use these method without modification. We follow the prescription proposed by Sch\"onhammer and Meden~\cite{rf:6}. Creation and annihilation operators of electrons can be described with boson operators as follows, 
\begin{eqnarray}
&& c_{l}=\frac{1}{2\pi}\int\nolimits_{0}^{2\pi}e^{-iul}\psi(u)du\ ,
\\ \nonumber \\
&& \psi^\dagger(u)\psi(v)=(\sum_{l=-\infty}^{N}e^{-il(u-v)}) \nonumber\\
&& \qquad \qquad \qquad 
\times   e^{-i(\phi^\dagger(u)-\phi^\dagger(v))}e^{-i(\phi(u)-\phi(v))}\ ,
\\ \nonumber \\
&&\phi(u)=-i\sum_{n=1}^{\infty}\frac{e^{inu}}{\sqrt{n}}b_{n}\ ,
\end{eqnarray} 
where $\psi(v)$ is an auxiliary field, $N$ is the number of electrons and $b_{n}$ is an operator which annihilates a boson with an angular momentum $n$. Substituting these equations into eq.(4) and expanding it up to quadratic order terms of boson operators, we can obtain a bosonized form of the Hamiltonian,
\begin{eqnarray}
H&=&\sum_{k} [k\left(\hbar\sqrt{\omega_{0}^2+\frac{1}{4}\omega_{c}^2}-\frac{1}{2}\hbar\omega_{c}\right)\nonumber\\
&+&\frac{1}{k}\sum_{j=N-k+1}^{N}\sum_{l=N+1}^{N+k}V_{k,j,l}\nonumber\\
&+&\frac{2}{k}\sum_{j=-\infty}^{N}\sum_{m=1}^{k}(V_{0,j,N+m}-V_{0,N-m+1,j})]b_{k}^\dagger b_{k}\nonumber\\
&+&o(b^3)+\cdot\cdot\cdot\ ,
\end{eqnarray}
here, the ground state is the vaccum of bosons. $k$ is defined as $k=L_{\rm {tot}}-L_{\rm {mim}}$, where $L_{\rm {mim}}$ is a total angular momentum of electrons which occupy the orbitals with the lowest values of angular momentum allowed by the Pauli principle. $V_{k,j,l}$ equals $0$ when $j$ or $l$ is negative. Higher order terms of bosons are less dominant when $N>>k$. The coefficients of third order terms of bosons are much smaller than those of quadratic order terms (less than 1/10 of harmonic terms) as will be described in detail later. We consider only quadratic order terms of bosons in this section.

Comparing the Hamiltonian $(8)$ with the bosonized form of TL Hamiltonian, the first term in the coefficient of $b_{k}^\dagger b_{k}$ has the same form as that of TL Hamiltonian. The second term is also almost the same and exactly corresponds to that of TL hamiltonian when $j,l$-dependence does not exist. The third term is the most important in this case because this term causes the edge reconstruction. It is always negative, so there is a possibility that the excitation energy also becomes negative. This term is does not exist without $j,l$-dependence.

Additionally, the Hamiltonian (8) has several remarkable features. It includes the terms of $V_{0,j,l}$ (angular momentum transfer=0) and $N$ which are ordinary classified into the ground state in TL model.

Fig.1 shows the excitation spectra for several values of magnetic field $B$ when $N=50$ and $\hbar\omega_{0}=2.3$[meV]. 
\begin{figure}[htbp]
\includegraphics[scale=0.54]{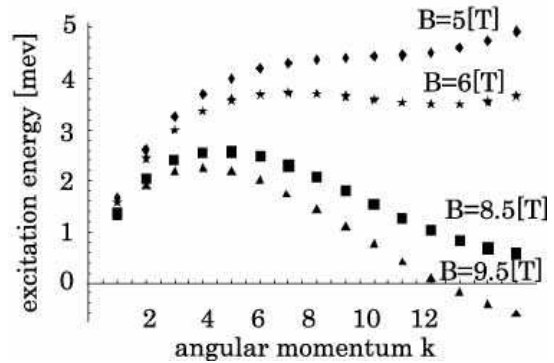}
\caption{\label{fig:epsart}
Excitation energy spectra for several values of the magnetic field. $\hbar\omega_{0}=2.3$[meV] and $N=50$}
\end{figure}
\begin{figure}[htbp]
\includegraphics[scale=0.54]{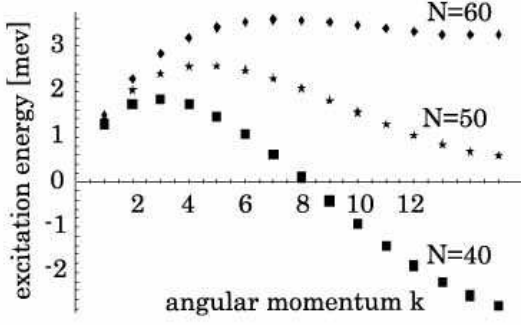}
\caption{\label{fig:epsart}
Excitation energy spectra for several values of the number of electrons. $\hbar\omega_{0}=2.3$[meV] and $B=8.5$[T]}
\end{figure}
\begin{figure}[htbp]
\includegraphics[scale=0.54]{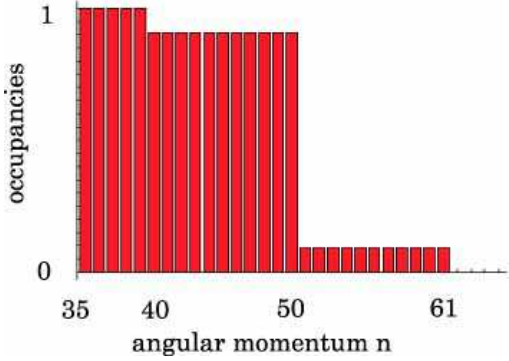}
\caption{\label{fig:epsart}
Occupation coefficients of single particle orbitals $<c_{n}^\dagger c_{n}>$ considering up to quadratic order terms of bosons in eq.(8). This graph shows the occupancies when a boson with an angular momentum $k=11$ is excited. The range of a reconstructed part is from $n=N-k+1$ to $n=N+k$ and expressed as a simple step function.}
\end{figure}
We take the number of electrons sufficiently large to be able to use the bosonization method. This figure shows that the edge reconstruction occurs with increase of magnetic field. As it increases, the difference between kinetic energies of neighboring single particle orbitals decreases, and the energy of the coulomb interaction increases. Those are the causes of the edge reconstruction. In our work, the edge reconstruction occurs between $B=8.5$[T] and $B=9.5$[T]. Moreover we can observe a characteristic minimum of the energy at $k=13$ for $B=6$[T], such minimum is also confirmed in the numerical diagonalization and referred to as the edge magneto-roton~\cite{rf:4}.

Fig.2 shows the energy spectra for several values of the number of electrons when $\hbar\omega_{0}=2.3$[meV] and $B=8.5$[T]. We confirm that the edge reconstruction occurs with decrease of the number of electron $N$, due to the increase of the energy of Coulomb interaction. The dependence of the energy spectra on the number of the electrons is also investigated with the numerical diagonalization~\cite{rf:4}, and the behaviors of the energy spectra in this paper is consistent with those in the numerical analysis. 

Next, we calculate the occupation coefficients $<c_{n}^\dagger c_{n}>$ of single particle orbitals which can be calculated with eq.(5)$\sim$(7),
\begin{eqnarray}
<k|c_{n}^\dagger c_{n}|k>&=&\sum_{l=0}^{\infty}[(1-\frac{2}{k})\delta_{n-N,-l}
\nonumber\\
&+&\frac{1}{k}(\delta_{n-N,-l-k}+\delta_{n-N,-l+k})]\ ,
\end{eqnarray}
 where $|k>$ is the state in which an elementary excitation of bosons with angular momentum $k$ is excited. We cannot get appropriate occupation coefficients as long as considering up to quadratic order terms of bosons. Fig.3 shows an occupation coefficient and we cannot confirm the configuration that a smaller droplet is surrounded by a ring, which is one of the important feature of the edge reconstruction and confirmed in the numerical analysis. So we have to consider higher order terms of bosons to obtain such an appropriate configuration.

\section{Third order terms of bosons}
In this section, we discuss the third order terms of bosons in Eq (8). It can be calculated as follows,
\begin{widetext}
\begin{eqnarray}
&&H_{3rd} = \sum_{n_{1}=n_{2}+n_{3}}\frac{1}{\sqrt{n_{1}n_{2}n_{3}}}
[(\sum_{j=N+1}^{N+n_{2}}\sum_{l=N+1}^{N+n_{3}}-\sum_{j=N-n_{1}+1}^{N-n_{3}}\sum_{l=N+1}^{N+n_{3}})V_{n_{3},j,l} \nonumber \\
&&+(\sum_{j=-\infty}^{N}\sum_{l=N+n_{3}+1}^{N+n_{1}}-\sum_{j=-\infty}^{N}\sum_{l=N+1}^{N+n_{2}}+\sum_{j=-\infty}^{N}\sum_{l=N-n_{1}+1}^{N-n_{3}}-\sum_{j=-\infty}^{N}\sum_{l=N-n_{2}+1}^{N})V_{0,j,l}
 \nonumber \\
&&+\frac{1}{2}(\sum_{j=N-n_{2}+1}^{N}\sum_{l=N+1}^{N+n_{1}}-\sum_{j=N-n_{1}+1}^{N-n_{3}}\sum_{l=N+1}^{N+n_{1}})V_{n_{1},j,l}](b_{n_{3}}^\dagger b_{n_{2}}^\dagger b_{n_{1}}+b_{n_{1}}^\dagger b_{n_{2}} b_{n_{3}})\ .
 \nonumber \\
\end{eqnarray}
\end{widetext}
As mentioned previously, most of the coefficients of third order trems are small enough compared to those of quadratic order terms, so we treat third order terms with a perturbation theory.
 we take into account up to second-order perturbations with respect to energies and up to first-order perturbations with respect to energy eigenstates. Both of them include processes that a boson with angular momentum $k$ splits into two particles with angular momentum $k-n$ and $n$ as an intermediate state,
\begin{eqnarray}
E_{k}^{(2)}&=&\sum_{n=1}^{[k/2]}\frac{|<k-n, n|H_{3rd}|k>|^2}{E_{k}^{(0)}-(E_{n}^{(0)}+E_{k-n}^{(0)})}\ , \\ \nonumber \\
|k>_{\rm {pert.}}&=&|k>\nonumber\\
&+&\sum_{n=1}^{[k/2]}|n, k-n>\frac{<k-n, n|H_{3rd}|k>}{E_{k}^{(0)}-(E_{n}^{(0)}+E_{k-n}^{(0)})}\ . \nonumber\\
\end{eqnarray}
Where $|k-n,n>$ is the state that bosons with angular momentum $k-n$ and $n$ are excited and $|k>_{\rm {pert.}}$ is the state including the effect of third order terms as a perturbation. $[k/2]$ means the integer part of $k/2$. $E_{k}^{(0)}$ is the unperturbed part of energy i.e., quadratic order terms in eq.(8). $E_{k}^{(2)}$ is the second-order perturbation with respect to $H_{3rd}$. The first-order terms of the perturbation does not appear.

Fig.4 shows excitation spectra when $\hbar\omega_{0}=2.3$[meV], $B=9.5$[T] and $N=50$. The spectrum A is unperturbed one and the other spectrum B includes the effect of perturbations. We can confirm the decreases of energies due to the second-order perturbation. As an angular momentum $k$ increases, a deviation from the unperturbed spectrum also increases because the number of the possible combinations of $k-n$ and $n$ increases.

Figs.5 and 6 show the occupation coefficients $<c_{n}^\dagger c_{n}>$ calculated with a perturbed state, eq.(12). The function form of $<c_{n}^\dagger c_{n}>$ is so complicated, composed  of such a great number of Kronecker deltas that we don't transcribe it on this paper. These graphs show the occupancies of the ground state with total angular momentum
\begin{figure}[htbp]
\includegraphics[scale=0.54]{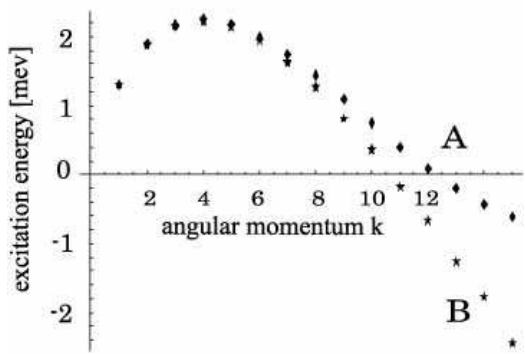}
\caption{\label{fig:epsart}
Excitation energy spectra when $\hbar\omega_{0}=2.3$[meV], $B=9.5$[T] and $N=50$. The spectrum A is unperturbed and B is perturbed.}
\end{figure}
\begin{figure}[htbp]
\includegraphics[scale=0.54]{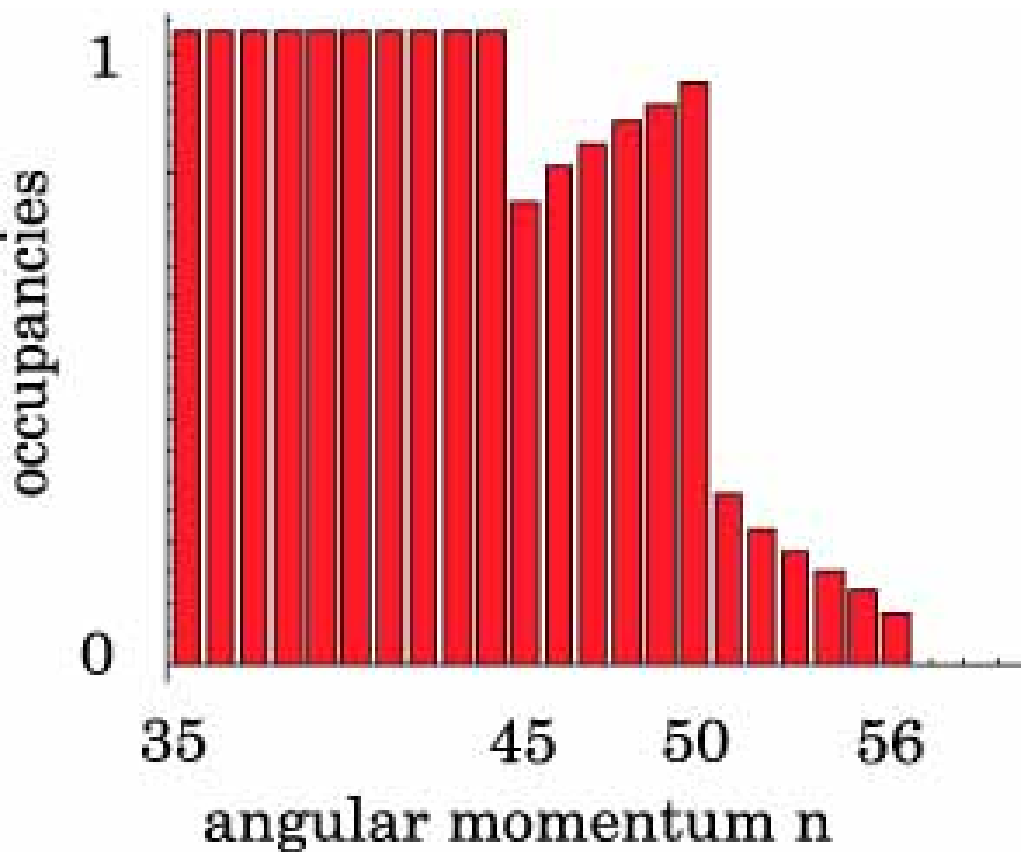}
\caption{\label{fig:epsart}
Occupation coefficients $<c_{n}^\dagger c_{n}>$ including the effect of third order terms of bosons when $N=50$ and total angular momentum $L_{\bf {tot}}=56$ $(k=6)$.}
\end{figure}
\begin{figure}[htbp]
\includegraphics[scale=0.54]{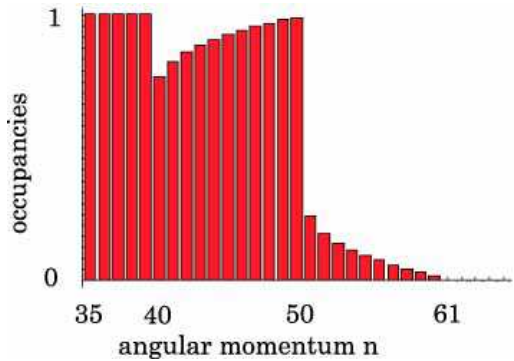}
\caption{\label{fig:epsart}
Occupation coefficients $<c_{n}^\dagger c_{n}>$ including the effect of third order terms of bosons when $N=50$ and total angular momentum $L_{\bf {tot}}=61$ $(k=11)$.}
\end{figure}
 $L_{\bf {tot}}=61$ $(k=11)$ and $L_{\bf {tot}}=56$ $(k=6)$ when $N=50$. Occupancies are fixed independent of confinement potential and magnetic field because eq.(12) is independent of $\omega_{0}$ and $B$. We can confirm the configurations from both graphs that a smaller droplet is surrounded by a ring. These configurations shows the tendency that the reconstructed part escapes further from the center of the dot than in the case of Figure 3. The more vacancy exists between the edge and the center, the more Coulomb interaction reduces. This means that the third order terms of bosons make a contribution to the reduction of Coulomb interaction energy.

\section{Summary}
In summary, we have developed a bosonization scheme to enable to treat such Hamiltonians as $(4)$. 
Using this method, we have obtained the excitation spectra analytically for several values of $B$ and $N$ and observed the edge reconstruction. We have also observed the appearance of the edge magneto-roton.

In addition to these results, we have examined the effect of the third order terms of bosons and obtain appropriate occupancies. The third order terms of bosons produce a ring at the edge of the dot and reduce the energy of Coulonb interaction. 

 Such behaviors as above agree well with the numerical analysis and we can say that the edge reconstruction has been recognized in a more analytical form, with a bosonization method.
\begin{acknowledgments}
We thank K. Kamide and N. Yokoshi for significant comments and discussions.
 This work is partly supported by The 21th Century COE Program at Waseda University from the Ministry of Education, Sports, Culture, Science and Technology of Japan.
\end{acknowledgments}
\section{Appendix}
The matrix element of Coulomb interaction $V_{i,j,k}$ is written as follows,
\begin{widetext}
\begin{eqnarray}
&&V_{k,j,l}=\frac{e^2}{2\pi \epsilon l_{0}}\sum_{l_{1}=0}^{j}\sum_{l_{2}=0}^{l}
\sum_{l_{3}=0}^{j+k}(-1)^{(-l_{1}+l_{3}-k)}\left(\frac{1}{2}\right)^{(j+l+\frac{5}{2})}\sqrt{\frac{1}{l!j!(j+k)!(l-k)}}
 \nonumber \\ \nonumber \\
&&\times\Gamma\left(l_{1}+l_{2}+1\right)\Gamma\left(\frac{2l+2j-2l_{1}-2l_{2}+1}{2}\right)
\left(
\begin{array}{cc}
j \\
l_{1} 
\end{array}
\right)
\left(
\begin{array}{cc}
l \\
l_{2} 
\end{array}
\right)
\left(
\begin{array}{cc}
j+k \\
l_{3} 
\end{array}
\right)
\left(
\begin{array}{cc}
l-k \\
l_{1}+l_{2}-l_{3} 
\end{array}
\right)\ .
 \nonumber \\
\end{eqnarray}
\end{widetext}
There are some other representations of $V_{k,j,l}$, e.g., in references~\cite{rf:2, rf:3} and which are consistent with ours.


\begin{thebibliography}{99}
\bibitem{rf:1} S. Tarucha {\it et al.}, Phys. Rev. Lett. {\bf 77}, 3613 (1996); L. P. Kouwenhoven {\it et al.}, Science {\bf 278}, 1788 (1997); D. G. Austing {\it et al.}, Phys. Rev. B {\bf 60}, 11514 (1999).
\bibitem{rf:2} Pawel Hawrylak, Solid State Commun. {\bf 88}, 475 (1993).
\bibitem{rf:3} Lucjan Jacak, Pawel Hawlylak and Arkadiusz W\"ojs, {\it Quantum Dots} (Springer, Berlin, 1997). 
\bibitem{rf:4} Pawel Hawrylak, Arkadiusz Wojs, and Jos\'e A. Brum, Phys. Rev. B {\bf 54}, 11397 (1996).
\bibitem{rf:5} D.C. Mattis and E. H. Lieb, J. Math. Phys. {\bf 6}, 304 (1965); A. Luther and I. Peschel, Phys. Rev. B {\bf 9}, 2911 (1974); F. D. M Haldane, J. Phys. C {\bf 14}, 2585 (1981).
\bibitem{rf:6} K. Sch\"onhammer and V. Meden, Am. J. Phys. {\bf 64}, 1168 (1996).
\bibitem{rf:7} H. Westfahl,Jr., A. H. Castro Neto and A. O. Caldeira, Phys. Rev. B {\bf 55} 7347 (1997).
\bibitem{rf:8} Paula Rojt, Yigal Meir and Assa Auerbach, Phys. Rev. Lett. {\bf 89}, 256401 (2002).
\bibitem{rf:9} Yogesh N. Joglekar, Hoang K. Nguyen and Ganpathy Murthy, Phys. Rev. B {\bf 68}, 035332 (2003).
\bibitem{rf:10} Xin Wan, Kun Yang and E. H. Rezayi Phys. Rev. Lett. {\bf 88}, 056802 (2002).
\bibitem{rf:11} Kun Yang, Phys. Rev. Lett. {\bf 91}, 036802 (2003).
\bibitem{rf:12} C. de .C. Chamon and X. G. Wen, Phys. Rev. B {\bf 49}, 8227 (1994)
\bibitem{rf:13} X.-G. Wen, Int. Mod. Phys. B {\bf 6}, 1711 (1992).
\end{thebibliography}
\end{document}